\documentclass[12pt]{article}
\usepackage{amsfonts}
\usepackage{amsmath}
\usepackage{amssymb}
\usepackage{graphicx}
\usepackage{color}
\usepackage[all, knot]{xy}
\usepackage{tikz}
\usepackage{array}

\usepackage{ulem}

\usepackage[utf8]{inputenc}
\usepackage{epstopdf}
\usepackage[footnotesize]{caption}
\usepackage{amsthm}
\usepackage{enumitem}
\usepackage{mathrsfs}

\usepackage[margin=3cm]{geometry}

\def \be {\begin{equation}}
\def \ee {\end{equation}}
\def \bea {\begin{eqnarray}}
\def \eea {\end{eqnarray}}
\def \nn {\nonumber}

\def \rr {\raise.35ex\hbox{\small $\prime$}\kern-.17em{\mbox{\large $\imath$}}}

\def \dels {\partial\kern-.6em /\kern.1em}
\def \As {{A\kern-.5em / \kern.5em}}
\def \Ds {D\kern-.7em / \kern.5em}

\def \ks {k\kern-.5em /}
\def \ls {l\kern-.5em /}

\newcommand{\ci}[1]{}

\newcommand{\ba}{\begin{eqnarray}}
\newcommand{\ea}{\end{eqnarray}}
\newcommand{\bal}{\begin{align}}
\newcommand{\eal}{\end{align}}
\newcommand{\bay}[1]{\left(\begin{array}{#1}}
\newcommand{\eay}{\end{array}\right)}

\newcommand{\ie}{\textit{i.e.}, }

\newcommand{\hide}[1]{}

\newlist{axioms}{enumerate}{2}
\setlist[axioms,1]{label=\textbf{A\arabic{axiomsi}.}, ref=A\arabic{axiomsi}}
\setlist[axioms,2]{label=\textbf{A\arabic{axiomsi}\rlap{\myEnumCounter{axiomsii}}.},%
                   ref=A\arabic{axiomsi}\myEnumCounter{axiomsii},%
                   align=parleft,%
                   leftmargin=0em,%
                   itemsep=1.4ex,%
                   before={\stepcounter{axiomsi}}}

  \usetikzlibrary{decorations.markings}

\begin{document}
\begin{titlepage}

\begin{center}

\textbf{\LARGE
Seiberg-Witten Map for D-branes in\\ 
Large R-R Field Background
\vskip.3cm
}
\vskip .5in
{\large
Chen-Te Ma$^{a, b, c, d}$ \footnote{e-mail address: yefgst@gmail.com}
\\
\vskip 1mm
}
{\sl
$^a$
Guangdong Provincial Key Laboratory of Nuclear Science,\\ 
Institute of Quantum Matter,
South China Normal University, Guangzhou 510006, Guangdong, China.
\\
$^b$
School of Physics and Telecommunication Engineering,\\
 South China Normal University, 
 Guangzhou 510006, Guangdong, China.
\\
$^c$
Guangdong-Hong Kong Joint Laboratory of Quantum Matter,\\
 Southern Nuclear Science Computing Center, 
South China Normal University, Guangzhou 510006, China. 
\\
$^d$ 
The Laboratory for Quantum Gravity and Strings,\\
 Department of Mathematics and Applied Mathematics,
University of Cape Town, Private Bag, Rondebosch 7700, South Africa.
}
\\
\vskip 1mm
\vspace{40pt}
\end{center}
\begin{abstract}
We obtain a Seiberg-Witten map for the gauge sector of multiple D$p$-branes in a large R-R $(p-1)$-form field background up to the first-order in the inverse R-R field background.
By applying the Seiberg-Witten map and then electromagnetic duality on the non-commutative D3-brane theory in the large R-R 2-form background, we find the expected commutative diagram of the Seiberg-Witten map and electromagnetic duality.
Extending the U(1) gauge group to the U($N$) gauge group, we obtain a commutative description of the D-branes in the large R-R field background. 
This construction is different from the known result.
\end{abstract}
\end{titlepage}

\section{Introduction}
\label{sec:1}
\noindent
String theory describes 1d object, string. 
A string can be either closed and forming a closed-loop, or open, \ie a segment with two endpoints. 
When a distance scale is much larger than the string scale, strings should behave like ordinary particles. 
The string states should determine all parameters.  
The endpoints of an open string lie in a D-brane.  
Requiring conformal invariance, the low-energy description of a D-brane is Dirac-Born-Infeld theory \cite{Abouelsaood:1986gd, Callan:1986bc}. 
Because open string theory has the non-commutative relation between the target spaces due to the endpoints, the D-brane can be studied in a non-commutative space \cite{Chu:1998qz}.
\\

\noindent 
The {\it non-commutative} and {\it commutative} descriptions are equivalent by the field redefinition, {\it Seiberg-Witten (SW) map} \cite{Seiberg:1999vs, Cornalba:1999ah, Okawa:1999cm, Asakawa:1999cu, Ishibashi:1999vi}. 
By assuming that the {\it SW map} and {\it gauge transformation} commute, the {\it SW map} can be found \cite{Seiberg:1999vs}. 
The Moyal product describes the non-commutative D-brane theory.
The non-commutativity parameter in the Moyal product introduces stringy or $\alpha^{\prime}$ correction. 
The parameter is inversely proportional to the Neveu-Schwarz–Neveu-Schwarz (NS-NS) field background. 
Hence the non-commutative gauge theory is useful in the study of brane theory.
\\

\noindent
People also extended the study of non-commutative gauge theory to the M-theory. 
The existence of M-theory was conjectured based on the duality.  
We still do not have a complete study in the M-theory, but it should contain membranes (M2-branes) and M5-branes. 
The entropy of coincident $N$ D-branes \cite{Tseytlin:1997csa} scales as $N^2$ at the large-$N$ limit, but that of multiple M5-branes scale as $N^3$. 
We expect that the M5-branes should be described by a new gauge formulation, not as the D-branes. 
Hence studying M5-branes should be useful for obtaining clues of the M-theory. 
An open membrane ending on an M5-brane (M2-M5 brane) forms an M5-brane. 
The non-commutative description of M5-brane theory in a large $C$-field background \cite{Ho:2008nn, Ho:2008ei} was constructed from infinitely many M2-branes or Bagger-Lambert-Gustavsson model \cite{Bagger:2006sk, Gustavsson:2007vu, Bagger:2007jr, Bagger:2007vi}. 
This M2-M5 brane theory is called Nambu-Poisson (NP) M5 theory. 
Because the NP M5-brane theory is in a large $C$-field background limit, the new gauge symmetry, the volume-preserving diffeomorphism, is described by the Nambu-Poisson bracket.
By the double dimensional reduction, which simultaneously compactifies directions in the target space and on an M5-brane, the NP M5 theory is consistently reduced to the non-commutative description of a D4-brane in a large NS-NS $B$-field background \cite{Ho:2008ve} and also the new non-commutative gauge theory of D4-brane in a large Ramond-Ramond (R-R) $C$-field background \cite{Ho:2011yr}. 
By the T-duality, the non-commutative D4-brane in a large $C$-field background can be extended to the D$p$-brane in a large $(p-1)$-form field background \cite{Ho:2013paa}. 
The Nambu-Poisson bracket was also extended to the $(p-1)$-bracket \cite{Ho:2013paa}. 
The non-commutative D3-brane in a large NS-NS background is S-dual to the non-commutative D3-brane in a large R-R background \cite{Ho:2013opa, Ho:2015mfa}. 
Hence the non-commutative D-brane theory in a large R-R field background is a successful construction. 
So far, the SW map and the extension to multiple D-branes are not known. 
The central question that we would like to address in this letter is the following: {\it What is the relation between the commutative and non-commutative descriptions in the large R-R field background?}    
\\

\noindent
In this letter, we obtain a solution of the SW map for the gauge sector of D-branes in the large R-R field background up to the first-order in the inverse R-R field background. 
The electromagnetic (EM) duality should be expected to {\it commute} with the SW map because the EM duality can be seen as a transformation for exchanging the electric and magnetic fields without changing a theory. 
We reduce the solution to the U(1) case and confirm that the solution can form a commutative diagram between the {\it EM duality} and {\it SW map}. 
By integrating out a 2-form gauge potential and changing the {\it gauge group}, the commutative description of the D4-brane can be extended to the multiple D4-branes. 
Finally, we discuss the difference between our description of the non-commutative multiple D4-branes in a large R-R $C$-field background and the known construction \cite{Ho:2013paa} from the {\it SW map}.

\section{SW Map for D-branes}
\label{sec:2}
\noindent
We first review the necessary contents of multiple D-branes in a large R-R field background \cite{Ho:2011yr} for solving the SW map \cite{Seiberg:1999vs}.

\subsection{Review of Multiple D$p$-branes}
\noindent
To review the multiple D-branes in a large R-R field background, we begin from more familiar U(1) gauge symmetry on a single D-brane. 
The scaling limit for the D$p$-brane gives a good approximation for such a R-R background \cite{Ho:2013paa}: 
\bea
l_s\sim\epsilon^{1/2}; \ 
g_s\sim\epsilon^{-1/2}; \
C_{\dot{\mu_1}\dot{\mu_2}\cdots\dot{\mu}_{p-1}}\sim\epsilon^0; \  
g_{\alpha\beta}\sim\epsilon^0; \ 
g_{\dot{\mu}\dot{\nu}}\sim\epsilon; \ 
\epsilon\rightarrow 0,
\eea
where $l_s\equiv(\alpha^{\prime})^{1/2}$ is the string length, $g_s$ is the string coupling constant, and $g_{\alpha\beta}$ and $g_{\dot{\mu}\dot{\nu}}$ are the spacetime metric. 
The $C_{\dot{1}\dot{2}\cdots\dot{p}-\dot{1}}$ is a large R-R $(p-1)$-form field background and only along the $\dot{\mu}=\dot{1}, \dot{2},\cdots ,\dot{p}-\dot{1}$ directions. 
Other directions of spacetime are labeled by $\alpha=0, 1$. 
The spacetime indices for all directions are labeled by $I=(\alpha, \dot{\mu})$. 
\\

\noindent
The gauge transformation for the gauge sector in the non-commutative D$p$-brane theory is \cite{Ho:2013paa}:
\bea
\hat{\delta}_{\hat{\Lambda}}\hat{b}^{\dot{\mu}}&=&\hat{\kappa}^{\dot{\mu}}
+g\hat{\kappa}^{\dot{\nu}}\partial_{\dot{\nu}}\hat{b}^{\dot{\mu}};
\nn\\
\hat{\delta}_{\hat{\Lambda}}\hat{a}_{\dot{\mu}}&=&\partial_{\dot{\mu}}\hat{\lambda}
+g(\hat{\kappa}^{\dot{\nu}}\partial_{\dot{\nu}}\hat{a}_{\dot{\mu}}
+\hat{a}_{\dot{\nu}}\partial_{\dot{\mu}}\hat{\kappa}^{\dot{\nu}});
\nn\\
\hat{\delta}_{\hat{\Lambda}}\hat{a}_{\alpha}&=&\partial_{\alpha}\hat{\lambda}
+g(\hat{\kappa}^{\dot{\nu}}\partial_{\dot{\nu}}\hat{a}_{\alpha}
+\hat{a}_{\dot{\nu}}\partial_{\alpha}\hat{\kappa}^{\dot{\nu}}),
\eea 
where the fields with $\ \hat{}\ $ denote the non-commutative description. 
A notation without the $\ \hat{}\ $ denotes the commutative description of fields later.
The coupling constant $g$ is defined as $g\equiv 1/C_{\dot{1}\dot{2}\cdots\dot{p}-\dot{1}}$.
The spacetime indices are raised or lowered by the flat metric $\eta_{IJ}\equiv\mathrm{diag}(-, +, +, \cdots, +)$.
The $\hat{\kappa}^{\dot{\mu}}$ and $\hat{\lambda}$ are gauge parameters,
which are associated with the gauge potentials $\hat{b}$ and $\hat{a}$ respectively. 
The $\hat{\kappa}^{\dot{\mu}}$ is divergenceless, $\partial_{\dot{\mu}}\hat{\kappa}^{\dot{\mu}}=0$, and it generates the volume-preserving diffeomorphism.
The covariant U(1) field strength is defined as the following \cite{Ho:2013paa}:
\bea
\hat{{\cal F}}_{\dot{\mu}\dot{\nu}}&\equiv&\frac{g^{p-3}}{(p-3)!}\epsilon_{\dot{\mu}\dot{\nu}\dot{\mu}_1\dot{\mu}_2\cdots\dot{\mu}_{p-3}}\{\hat{X}^{\dot{\mu}_1}, \hat{X}^{\dot{\mu}_2}, \cdots, \hat{X}^{\dot{\mu}_{p-3}}, \hat{a}_{\dot{\rho}}, y^{\dot{\rho}}\};
\nn\\
\hat{{\cal F}}_{\alpha\dot{\mu}}&\equiv& \hat{V}^{-1}{}_{\dot{\mu}}{}^{\dot{\nu}}(\hat{F}_{\alpha\dot{\nu}}+g\hat{F}_{\dot{\nu}\dot{\delta}}\hat{B}_{\alpha}{}^{\dot{\delta}});
\nn\\
\hat{{\cal F}}_{\alpha\beta}&\equiv&\hat{F}_{\alpha\beta}+g(-\hat{F}_{\alpha\dot{\mu}}\hat{B}_{\beta}{}^{\dot{\mu}}
-\hat{F}_{\dot{\mu}\beta}\hat{B}_{\alpha}{}^{\dot{\mu}})
+g^2\hat{F}_{\dot{\mu}\dot{\nu}}\hat{B}_{\alpha}{}^{\dot{\mu}}\hat{B}_{\beta}{}^{\dot{\nu}},
\eea 
where $\{f_1, f_2 ,\cdots, f_{p-1}\}\equiv \epsilon^{\dot{\mu}_1\dot{\mu}_2\cdots\dot{\mu}_{p-1}}(\partial_{\dot{\mu}_1}f_1)(\partial_{\dot{\mu}_2f_2})\cdots(\partial_{\dot{\mu}_{p-1}}f_{p-1})$. 
When $p$=1, the $(p-1)$-bracket is equivalent to the ordinary derivative and gives the length-preserving diffeomorphism. 
For $p=2$, it is called the Poisson bracket and gives the area-preserving diffeomorphism. 
The volume-preserving diffeomorphism is generated by the $(p-1)$-bracket. 
The $x^{\alpha}$ and $y^{\dot{\rho}}$ are the world-volume coordinates.
The $\hat{V}_{\dot{\nu}}{}^{\dot{\mu}}$ and $\hat{X}^{\dot{\mu}}$ are define as the following: 
\bea
\hat{V}_{\dot{\nu}}{}^{\dot{\mu}}\equiv\delta_{\dot{\nu}}{}^{\dot{\mu}}+g\partial_{\dot{\nu}}\hat{b}^{\dot{\mu}}; \qquad 
\hat{X}^{\dot{\mu}}\equiv\frac{y^{\dot{\mu}}}{g}+\hat{b}^{\dot{\mu}}.
\eea
 The $\hat{B}_{\alpha}{}^{\dot{\mu}}$ satisfies the following equation 
\bea
\hat{V}_{\dot{\mu}}{}^{\dot{\nu}}(\partial^{\alpha}\hat{b}_{\dot{\nu}}-\hat{V}^{\dot{\rho}}{}_{\dot{\nu}}\hat{B}^{\alpha}{}_{\dot{\rho}})
+\epsilon^{\alpha\beta}\hat{F}_{\beta\dot{\mu}}
+g\epsilon^{\alpha\beta}\hat{F}_{\dot{\mu}\dot{\nu}}\hat{B}_{\beta}{}^{\dot{\nu}}=0,
\eea
 where $\hat{F}_{IJ}\equiv \partial_I\hat{A}_J-\partial_J\hat{A}_I$ is the Abelian strength. 
\\

\noindent 
The gauge potential $\hat{a}_{\dot{\mu}}$ is necessarily T-dual to the scalar field upon compactification on a circle in the direction $\dot{\mu}$. 
The covariant property on the field strength $\hat{{\cal F}}_{\dot{\mu}\dot{\nu}}$ is not enough to restrict the form of the field strength \cite{Ho:2013paa}. 
Ones can show that the field strength is also covariant under the gauge transformation $\hat{{\cal G}}_{\dot{\mu}\dot{\nu}}\equiv
\hat{F}_{\dot{\mu}\dot{\nu}}
+g(\partial_{\dot{\sigma}}\hat{b}^{\dot{\sigma}}\hat{F}_{\dot{\mu}\dot{\nu}}
-\partial_{\dot{\mu}}\hat{b}^{\dot{\sigma}}\hat{F}_{\dot{\sigma}\dot{\nu}}
-\partial_{\dot{\nu}}\hat{b}^{\dot{\sigma}}\hat{F}_{\dot{\mu}\dot{\sigma}})$. 
However, the field strength cannot provide the covariant derivative acting on a scalar field under the T-duality \cite{Ho:2013paa}. 
Therefore, the field strength cannot provide a proper description to D-brane \cite{Ho:2013paa}. 
For the multiple D$p$-branes \cite{Ho:2011yr} that we will introduce, ones adopted the field strength $\hat{{\cal G}}_{\dot{\mu}\dot{\nu}}$. 
The proposal of the multiple D-branes \cite{Ho:2011yr} is necessary to be modified concerning the T-duality \cite{Ho:2013paa}. 
However, the field strength does not change when $p\le 4$ \cite{Ho:2013paa}. 
Hence the proposal is still not completely ruled out. 
\\

\noindent
Now we introduce the proposal of the multiple D$p$-branes. 
The gauge potential, $a_I$, contains different gauge groups in the proposal. 
Therefore, we will particularly specify the U(1) and SU($N$) gauge potentials without any confusion. 
The gauge transformation for the non-commutative multiple D$p$-branes theory in a large $(p-1)$-form field background is \cite{Ho:2011yr}: 
\bea
\hat{\delta}_{\hat{\Lambda}}\hat{b}^{\dot{\mu}}&=&\hat{\kappa}^{\dot{\mu}}
+g\hat{\kappa}^{\dot{\nu}}\partial_{\dot{\nu}}\hat{b}^{\dot{\mu}};
\nn\\
\hat{\delta}_{\hat{\Lambda}}\hat{a}_{\dot{\mu}}&=&\partial_{\dot{\mu}}\hat{\lambda}+
i\lbrack\hat{\lambda}, \hat{a}_{\dot{\mu}}\rbrack
+g(\hat{\kappa}^{\dot{\nu}}\partial_{\dot{\nu}}\hat{a}_{\dot{\mu}}
+\hat{a}_{\dot{\nu}}\partial_{\dot{\mu}}\hat{\kappa}^{\dot{\nu}});
\nn\\
\hat{\delta}_{\hat{\Lambda}}\hat{a}_{\alpha}&=&\partial_{\alpha}\hat{\lambda}+
i\lbrack\hat{\lambda}, \hat{a}_{\alpha}\rbrack
+g(\hat{\kappa}^{\dot{\nu}}\partial_{\dot{\nu}}\hat{a}_{\alpha}
+\hat{a}_{\dot{\nu}}\partial_{\alpha}\hat{\kappa}^{\dot{\nu}}).
\eea 
The gauge potential $\hat{b}$ takes the value of U(1), and the gauge potential $\hat{a}$ takes the value of U($N$). 
The commutator is defined as $\lbrack{\cal O}_1, {\cal O}_2\rbrack\equiv {\cal O}_1^c{\cal O}_2^d(T^cT^d-T^dT^c)$, where $T^c$ is the generator of a Lie algebra.
The indices of a Lie algebra are labeled by $c$ and $d$. 
Although the field strength is necessary to be modified, the gauge transformation is not. 
\\

\noindent
The gauge transformation is non-trivial because the covariant field strength for the non-Abelian group U($N$) can be generalized in a similar way from the U(1) case \cite{Ho:2011yr}: 
\bea
\hat{{\cal F}}_{\dot{\mu}\dot{\nu}}&\equiv& \hat{F}_{\dot{\mu}\dot{\nu}}
+g(\partial_{\dot{\sigma}}\hat{b}^{\dot{\sigma}}\hat{F}_{\dot{\mu}\dot{\nu}}
-\partial_{\dot{\mu}}\hat{b}^{\dot{\sigma}}\hat{F}_{\dot{\sigma}\dot{\nu}}
-\partial_{\dot{\nu}}\hat{b}^{\dot{\sigma}}\hat{F}_{\dot{\mu}\dot{\sigma}});
\nn\\
\hat{{\cal F}}_{\alpha\dot{\mu}}&\equiv& (\hat{V}^{-1})_{\dot{\mu}}{}^{\dot{\nu}}(\hat{F}_{\alpha\dot{\nu}}+g\hat{F}_{\dot{\nu}\dot{\delta}}\hat{B}_{\alpha}{}^{\dot{\delta}});
\nn\\
\hat{{\cal F}}_{\alpha\beta}&\equiv&\hat{F}_{\alpha\beta}+
g(-\hat{F}_{\alpha\dot{\mu}}\hat{B}_{\beta}{}^{\dot{\mu}}-\hat{F}_{\dot{\mu}\beta}\hat{B}_{\alpha}{}^{\dot{\mu}})+
g^2\hat{F}_{\dot{\mu}\dot{\nu}}\hat{B}_{\alpha}{}^{\dot{\mu}}\hat{B}_{\beta}{}^{\dot{\nu}},
\eea
where $\hat{B}_{\alpha}{}^{\dot{\mu}}$ takes the value of U(1), and $\hat{F}_{IJ}\equiv\partial_I\hat{a}_J-\partial_J\hat{a}_I-i\lbrack \hat{a}_I, \hat{a}_J\rbrack$. 
The $\hat{B}_{\alpha}{}^{\dot{\mu}}$ satisfies the following equation 
\bea
\hat{V}_{\dot{\mu}}{}^{\dot{\nu}}(\partial^{\alpha}\hat{b}_{\dot{\nu}}-\hat{V}^{\dot{\rho}}{}_{\dot{\nu}}\hat{B}^{\alpha}{}_{\dot{\rho}})
+\epsilon^{\alpha\beta}\hat{F}^{\mathrm{U}(1)}_{\beta\dot{\mu}}
+g\epsilon^{\alpha\beta}\hat{F}^{\mathrm{U}(1)}_{\dot{\mu}\dot{\nu}}\hat{B}_{\beta}{}^{\dot{\nu}}=0,
\eea
 where $\hat{F}_{IJ}^{\mathrm{U}(1)}$ is the Abelian strength. 
 As we mentioned, the covariant field strengths only possibly give a suitable description for $p\le 4$.
Obtaining a non-Abelian gauge theory from the Abelian gauge theory is necessary to introduce the non-Abelian gauge group to all gauge potentials before. 
This study has not done in the known construction of D-branes in a large R-R field background (the gauge group of $\hat{b}$ field remains U(1)) \cite{Ho:2011yr}. 
It should be interesting to solve the SW map and study the commutative description.

\subsection{SW Map for U($N$)}
\noindent
The SW map \cite{Seiberg:1999vs} commutes with the gauge transformation, and it satisfies the mathematical relations:
\bea
\hat{b}^{\dot{\mu}}(b+\delta_{\Lambda}b)-\hat{b}^{\dot{\mu}}(b)=\hat{\delta}_{\hat{\Lambda}}\hat{b}^{\dot{\mu}}; \qquad  
\hat{a}_{I}(a+\delta_{\Lambda}a, b+\delta_{\Lambda}b)-\hat{a}_{I}(a, b)=\hat{\delta}_{\hat{\Lambda}}\hat{a}_{I}.
\eea 
Up to the first-order in $g$, we obtain the solution for the multiple D-branes in the large R-R field background: 
\bea
\hat{b}^{\dot{\mu}}(b)&=&b^{\dot{\mu}}+\frac{g}{2}b^{\dot{\nu}}\partial_{\dot{\nu}}b^{\dot{\mu}}
+\frac{g}{2}b^{\dot{\mu}}\partial_{\dot{\nu}}b^{\dot{\nu}}+{\cal O}(g^2); 
\nn\\
\hat{a}_{I}(a, b)&=&a_{I}+g(b^{\dot{\rho}}\partial_{\dot{\rho}}a_{I}+a_{\dot{\rho}}\partial_{I}b^{\dot{\rho}})+{\cal O}(g^2);
\nn\\ 
\hat{\kappa}^{\dot{\mu}}&=&\kappa^{\dot{\mu}}
+\frac{g}{2}b^{\dot{\nu}}\partial_{\dot{\nu}}\kappa^{\dot{\mu}}
+\frac{g}{2}(\partial_{\dot{\nu}}b^{\dot{\nu}})\kappa^{\dot{\mu}}
-\frac{g}{2}(\partial_{\dot{\nu}}b^{\dot{\mu}})\kappa^{\dot{\nu}}
+{\cal O}(g^2);
\nn\\ 
\hat{\lambda}&=&\lambda+gb^{\dot{\rho}}\partial_{\dot{\rho}}\lambda+{\cal O}(g^2).
\eea 
Because the gauge group of $\hat{b}^{\dot{\mu}}$ and $\hat{\kappa}^{\dot{\mu}}$ remains U(1), the solution of the SW map remains the same as in the NP M5 brane theory \cite{Ho:2008ve}. 
The gauge group of $\hat{a}_I$ and $\hat{\lambda}$ is U($N$).
The gauge potential $\hat{a}^{\mathrm{U}(1)}_{\alpha}$ cannot come up without a complicated dual \cite{Ho:2011yr}, and the solution of the $\hat{\lambda}^{\mathrm{U}(1)}$ cannot be derived directly from the NP M5-brane theory \cite{Ho:2008nn}. 
The solution is consistent with the dimensional reduction from the NP M5-brane theory. 
It is also consistent with the generalization of the single D-brane. 
Hence the solution of the SW map is non-trivial. 
We will check the solution from the expected relation on the D3-brane.

\section{EM Duality and SW Map}
\label{sec:3}
\noindent
The non-commutative D3-brane in a large R-R field background \cite{Ho:2013paa} is electromagnetic dual to the 
non-commutative D3-brane in a large NS-NS field background \cite{Ho:2013opa, Ho:2015mfa}. 
The non-commutative D3-brane theory in a large NS-NS field background is also dual to the commutative D3-brane theory in the large NS-NS field background by the SW map \cite{Seiberg:1999vs}.
Therefore, it is only necessary to exchange the ordering of the EM duality and SW map. 
We then can conclude that the EM duality and SW map need to form a commutative diagram. 
\\

\noindent
Because no people solved the SW map for the D-brane theory in the large R-R field background and showed the commutative description, we first write down the commutative description of the D4-brane in the large R-R field background and begin from the action \cite{Ho:2011yr}
\bea
&&S_{NRR4}
\nn\\
&=&\int d^2xd^3\dot{x}\ \bigg(-\frac{1}{2}\hat{{\cal H}}_{\dot{1}\dot{2}\dot{3}}\hat{{\cal H}}^{\dot{1}\dot{2}\dot{3}}
-\frac{1}{4}\hat{{\cal F}}_{\dot{\nu}\dot{\rho}}\hat{{\cal F}}^{\dot{\nu}\dot{\rho}}
+\frac{1}{2}\hat{{\cal F}}_{\beta\dot{\mu}}\hat{{\cal F}}^{\beta\dot{\mu}}
+\frac{1}{2g}\epsilon^{\alpha\beta}\hat{{\cal F}}_{\alpha\beta}\bigg),
\eea
where 
\bea
\hat{{\cal H}}_{\dot{1}\dot{2}\dot{3}}&\equiv&
\partial_{\dot{\mu}}\hat{b}^{\dot{\mu}}
+\frac{g}{2}(\partial_{\dot{\nu}}\hat{b}^{\dot{\nu}}\partial_{\dot{\rho}}\hat{b}^{\dot{\rho}}
-\partial_{\dot{\nu}}\hat{b}^{\dot{\rho}}\partial_{\dot{\rho}}\hat{b}^{\dot{\nu}})
+g^2\{\hat{b}^{\dot{1}}, \hat{b}^{\dot{2}}, \hat{b}^{\dot{3}}\};
\nn\\
\{ {\cal O}_1, {\cal O}_2, {\cal O}_3\}&\equiv&\epsilon^{\dot{\mu}\dot{\nu}\dot{\rho}}\partial_{\dot{\mu}} {\cal O}_1\partial_{\dot{\nu}}{\cal O}_2\partial_{\dot{\rho}}{\cal O}_3.
\eea
This action was derived from the NP M5 action \cite{Ho:2008nn} through the double dimensional reduction on the $x^2$ direction directly \cite{Ho:2011yr}.
\\

\noindent
Applying the SW map, we rewrite the covariant field strength in terms of the commutative description of fields:
\bea
\hat{{\cal H}}_{\dot{1}\dot{2}\dot{3}}
&=&\partial_{\dot{\mu}}b^{\dot{\mu}}
+g(\partial_{\dot{\mu}}b^{\dot{\mu}}\partial_{\dot{\nu}}b^{\dot{\nu}}+b^{\dot{\mu}}\partial_{\dot{\mu}}\partial_{\dot{\nu}}b^{\dot{\nu}})
+{\cal O}(g^2);
\nn\\
\hat{{\cal F}}_{\dot{\mu}\dot{\nu}}
&=&
F_{\dot{\mu}\dot{\nu}}+g(b^{\dot{\rho}}\partial_{\dot{\rho}}F_{\dot{\mu}\dot{\nu}}+\partial_{\dot{\sigma}}b^{\dot{\sigma}}F_{\dot{\mu}\dot{\nu}})+{\cal O}(g^2);
\nn\\
\hat{{\cal F}}_{\alpha\dot{\mu}}
&=&
F_{\alpha\dot{\mu}}+g(b^{\dot{\nu}}\partial_{\dot{\nu}}F_{\alpha\dot{\mu}}+\epsilon_{\alpha\beta}F_{\dot{\mu}\dot{\sigma}}F^{\beta\dot{\sigma}})
+{\cal O}(g^2);
\nn
\eea
\bea
&&\hat{{\cal F}}_{\alpha\beta}
\nn\\
&=&
\hat{F}_{\alpha\beta}
\nn\\
&&
+g(-F_{\alpha\dot{\mu}}\partial_{\beta}b^{\dot{\mu}}
-\epsilon_{\beta\gamma}F_{\alpha\dot{\mu}}F^{\gamma\dot{\mu}}
-F_{\dot{\mu}\beta}\partial_{\alpha}b^{\dot{\mu}}
-\epsilon_{\alpha\gamma}F_{\dot{\mu}\beta}F^{\gamma\dot{\mu}})
\nn\\
&&
+g^2(
-F_{\alpha\dot{\mu}}F_{\beta\dot{\sigma}}F^{\dot{\mu}\dot{\sigma}}
-F_{\dot{\mu}\beta}F_{\alpha\dot{\sigma}}F^{\dot{\mu}\dot{\sigma}}
\nn\\
&&
+\epsilon_{\alpha\gamma}\epsilon_{\beta\delta}F_{\dot{\mu}\dot{\nu}}F^{\gamma\dot{\mu}}F^{\delta\dot{\nu}}
-b^{\dot{\nu}}\partial_{\dot{\nu}}F_{\alpha\dot{\mu}}\partial_{\beta}b^{\dot{\mu}}
-\frac{1}{2}F_{\alpha\dot{\mu}}\partial_{\beta}b^{\dot{\nu}}\partial_{\dot{\nu}}b^{\dot{\mu}}
-\frac{1}{2}F_{\alpha\dot{\mu}}b^{\dot{\nu}}\partial_{\beta}\partial_{\dot{\nu}}b^{\dot{\mu}}
\nn\\
&&
-\frac{1}{2}F_{\alpha\dot{\mu}}\partial_{\beta}b^{\dot{\mu}}\partial_{\dot{\nu}}b^{\dot{\nu}}
-\frac{1}{2}F_{\alpha\dot{\mu}}b^{\dot{\mu}}\partial_{\beta}\partial_{\dot{\nu}}b^{\dot{\nu}}
-\epsilon_{\beta\gamma}b^{\dot{\nu}}\partial_{\dot{\nu}}F_{\alpha\dot{\mu}}F^{\gamma\dot{\mu}}
\nn\\
&&
-\epsilon_{\beta\gamma}F_{\alpha\dot{\mu}}b^{\dot{\nu}}\partial_{\dot{\nu}}F^{\gamma\dot{\mu}}
-\partial_{\beta}b^{\dot{\nu}}F_{\dot{\mu}\dot{\nu}}\partial_{\alpha}b^{\dot{\mu}}
+b^{\dot{\nu}}\partial_{\dot{\nu}}F_{\beta\dot{\mu}}\partial_{\alpha}b^{\dot{\mu}}
\nn\\
&&
+\frac{1}{2}F_{\beta\dot{\mu}}\partial_{\alpha}b^{\dot{\nu}}\partial_{\dot{\nu}}b^{\dot{\mu}}
+\frac{1}{2}F_{\beta\dot{\mu}}b^{\dot{\nu}}\partial_{\alpha}\partial_{\dot{\nu}}b^{\dot{\mu}}
\nn\\
&&
+\frac{1}{2}F_{\beta\dot{\mu}}\partial_{\alpha}b^{\dot{\mu}}\partial_{\dot{\nu}}b^{\dot{\nu}}
+\frac{1}{2}F_{\beta\dot{\mu}}b^{\dot{\mu}}\partial_{\alpha}\partial_{\dot{\nu}}b^{\dot{\nu}}
+\epsilon_{\alpha\gamma}b^{\dot{\nu}}\partial_{\dot{\nu}}F_{\beta\dot{\mu}}F^{\gamma\dot{\mu}}
+
\epsilon_{\alpha\gamma}F_{\beta\dot{\mu}}b^{\dot{\nu}}\partial_{\dot{\nu}}F^{\gamma\dot{\mu}})
\nn\\
&&
+{\cal O}(g^3).
\eea
Note that the field strength $\hat{F}_{\alpha\beta}$ only appears in a boundary term. 
This term will not appear in our action. 
Hence we do not transform the $\hat{F}_{\alpha\beta}$. 
\\

\noindent
After substituting the new covariant field strength, we obtain the action:
\bea
&&S_{CRR4}
\nn\\
&=&
\int d^2xd^3\dot{x}\ \bigg\lbrack-\frac{1}{2}H_{\dot{1}\dot{2}\dot{3}}H_{\dot{1}\dot{2}\dot{3}}
-\frac{1}{2}\epsilon^{\alpha\beta}F_{\alpha\beta}H_{\dot{1}\dot{2}\dot{3}}
-\frac{1}{4}F_{\dot{\mu}\dot{\nu}}F^{\dot{\mu}\dot{\nu}}
-\frac{1}{2}F_{\alpha\dot{\mu}}F^{\alpha\dot{\mu}}
\nn\\
&&
+g\bigg(
-\frac{1}{2}H_{\dot{1}\dot{2}\dot{3}}H_{\dot{1}\dot{2}\dot{3}}H_{\dot{1}\dot{2}\dot{3}}
-\frac{1}{4}H_{\dot{1}\dot{2}\dot{3}}F^{\dot{\mu}\dot{\nu}}F_{\dot{\mu}\dot{\nu}}
+\frac{1}{2}H_{\dot{1}\dot{2}\dot{3}}F^{\alpha\dot{\mu}}F_{\alpha\dot{\mu}}
\nn\\
&&
-\frac{1}{2}\epsilon_{\alpha\beta}F^{\alpha\dot{\mu}}F^{\beta\dot{\sigma}}F_{\dot{\mu}\dot{\sigma}}\bigg)\bigg\rbrack
\nn\\
&=&
S_{NRR4}+{\cal O}(g^2),
\eea
which is up to a total derivative term.
The $S_{CRR4}$ describes the commutative D4-brane in a large R-R field background up to the first-order in $g$. 
We can find that the action can be rewritten in terms of the Abelian field strength, as expected from the gauge transformation of the commutative description. 
In the end, we integrate out the $b$ field under the gauge fixing $\epsilon^{\dot{\mu}\dot{\nu}\dot{\lambda}}\partial_{\dot{\mu}}b_{\dot{\nu}}=0$, which is equivalent to using $H_{\dot{1}\dot{2}\dot{3}}=-F_{01}+{\cal O}(g)$ \cite{Ho:2011yr}. 
The first-order correction for this integration does not affect the first-order correction of the below action:
\bea
&&S_{CRR4a}
\nn\\
&=&\int d^2xd^3\dot{x}\ \bigg\lbrack\bigg(-\frac{1}{4}F_{IJ}F^{IJ}
\nn\\
&&+g\bigg(
\frac{1}{2}F_{01}F_{01}F_{01}+\frac{1}{4}F_{01}F^{\dot{\mu}\dot{\nu}}F_{\dot{\mu}\dot{\nu}}
-\frac{1}{2}F_{01}F^{\alpha\dot{\mu}}F_{\alpha\dot{\mu}}
-\frac{1}{2}\epsilon_{\alpha\beta}F^{\alpha\dot{\mu}}F^{\beta\dot{\nu}}F_{\dot{\mu}\dot{\nu}}
\bigg)\bigg\rbrack.
\nn\\
\eea   
Now the degrees of freedom for the gauge potential $b^{\dot{\mu}}$ is replaced by the gauge potential $a_{\alpha}$. 
The leading-order is the standard Abelian gauge theory. 
\\

\noindent
Performing the dimensional reduction on the $x^{\dot{3}}$ direction on the D4-brane in the large R-R field background, we obtain the D3-brane theory in the large R-R field background. 
The action for the gauge sector is the same, except that the range of the index is changed ($\dot{\mu}=\dot{1}, \dot{2}, \dot{3}\rightarrow\dot{\mu}=\dot{1}, \dot{2}$). 
Note that exchanging the ordering of the SW map and dimensional reduction also reaches the same action.
\\

\noindent
Now we perform the EM duality. 
We promote the $F$ to an un-constrained field by introducing a new term $(1/2)\tilde{G}_{IJ}F^{IJ}$ to the D3-brane theory as in the below
\bea
&&S_{EM1}
\nn\\
&=&\int d^2xd^2\dot{x}\ \bigg\lbrack-\frac{1}{4}F_{IJ}F^{IJ}
\nn\\
&&+g\bigg(
\frac{1}{2}F_{01}F_{01}F_{01}+\frac{1}{4}F_{01}F^{\dot{\mu}\dot{\nu}}F_{\dot{\mu}\dot{\nu}}
-\frac{1}{2}F_{01}F^{\alpha\dot{\mu}}F_{\alpha\dot{\mu}}
-\frac{1}{2}\epsilon_{\alpha\beta}F^{\alpha\dot{\mu}}F^{\beta\dot{\nu}}F_{\dot{\mu}\dot{\nu}}
\bigg)
\nn\\
&&
+\frac{1}{2}\tilde{G}_{IJ}F^{IJ}\bigg\rbrack,
\eea  
where 
\bea
\tilde{G}_{IJ}\equiv\frac{1}{2}\epsilon_{IJKL}G^{KL}; \qquad G_{IJ}\equiv \partial_{I}\bar{A}_{J}-\partial_{J}\bar{A}_{I}.
\eea
We can integrate out $\bar{A}$, which is equivalent to using $dF=0$ or rather locally equivalent to $F=dA$. 
Hence this goes back to the D3-brane theory in the large R-R field background, 
and the new term does not change the theory. 
\\

\noindent
Now we integrate out $F$ to obtain the EM dual description, and it is equivalent to varying $F$ up to the first-order in $g$: 
\bea
0&=&-F_{01}+g\bigg(\frac{3}{2}F_{01}F_{01}-\frac{1}{2}F^{\alpha\dot{\mu}}F_{\alpha\dot{\mu}}\bigg)+\tilde{G}_{01}+{\cal O}(g^2); 
\nn\\
0&=&-F_{\dot{\mu}\dot{\nu}}+g(F_{01}F_{\dot{\mu}\dot{\nu}}-\epsilon^{\alpha\beta}F_{\alpha\dot{\mu}}F_{\beta\dot{\nu}})
+\tilde{G}_{\dot{\mu}\dot{\nu}}+{\cal O}(g^2); 
\nn\\ 
0&=&-F_{\alpha\dot{\mu}}+g(-F_{01}F_{\alpha\dot{\mu}}-\epsilon_{\alpha\beta}F^{\beta\dot{\nu}}F_{\dot{\mu}\dot{\nu}})
+\tilde{G}_{\alpha\dot{\mu}}+{\cal O}(g^2).
\eea
Up to the first-order in $g$, we can rewrite the $F$ in terms of the $\tilde{G}$: 
\bea
F_{01}&=&\tilde{G}_{01}+g\bigg(\frac{3}{2}\tilde{G}_{01}\tilde{G}_{01}-\frac{1}{2}\tilde{G}^{\alpha\dot{\mu}}\tilde{G}_{\alpha\dot{\mu}}\bigg)+{\cal O}(g^2); 
\nn\\ 
F_{\dot{\mu}\dot{\nu}}&=&\tilde{G}_{\dot{\mu}\dot{\nu}}
+g(\tilde{G}_{01}\tilde{G}_{\dot{\mu}\dot{\nu}}-\epsilon^{\alpha\beta}\tilde{G}_{\alpha\dot{\mu}}\tilde{G}_{\beta\dot{\nu}})+{\cal O}(g^2);
\nn\\ 
F_{\alpha\dot{\mu}}&=&\tilde{G}_{\alpha\dot{\mu}}+g(-\tilde{G}_{01}\tilde{G}_{\alpha\dot{\mu}}-\epsilon_{\alpha\beta}\tilde{G}^{\beta\dot{\nu}}\tilde{G}_{\dot{\mu}\dot{\nu}})+{\cal O}(g^2).
\eea
After substituting the above relation to the action $S_{EM1}$, the EM dual description is obtained as the following
\bea
&&S_{EM}
\nn\\
&=&\int d^2x d^2\dot{x}\ \bigg\lbrack-\frac{1}{4}G_{IJ}G^{IJ}
\nn\\
&&+g\bigg(-\frac{1}{2}G_{\dot{1}\dot{2}}G_{\dot{1}\dot{2}}G_{\dot{1}\dot{2}}
-\frac{1}{2}G_{\dot{1}\dot{2}}G_{01}G_{01}
-\frac{1}{2}G_{\dot{1}\dot{2}}G^{\alpha\dot{\mu}}G_{\alpha\dot{\mu}}
\nn\\
&&
-\frac{1}{2}\epsilon_{\alpha\beta\dot{\mu}\dot{\nu}}G^{\alpha\dot{\mu}}G^{\beta\dot{\nu}}G_{01}\bigg)\bigg\rbrack.
\eea
\\

\noindent
Now we introduce the D3-brane theory in the NS-NS field background and then consider the large NS-NS field background limit to compare to the $S_{EM}$. 
The non-commutative D3-brane theory in the NS-NS field background is described by the action \cite{Seiberg:1999vs}
\bea
S_{NNSNS3}=-\frac{1}{4}\int d^2xd^2\dot{x}\ \hat{F}_{IJ}*\hat{F}^{IJ},
\eea
where 
\bea
\hat{F}_{IJ}&\equiv&\partial_{I}\hat{A}_{J}-\partial_{J}\hat{A}_{I}+\lbrack\hat{A}_{I}, \hat{A}_{J}\rbrack_*;
\nn\\ 
{\cal O}_1*{\cal O}_2&\equiv& {\cal O}_1\exp\bigg(\frac{\theta^{IJ}}{2} \overleftarrow{\partial}_{I}\overrightarrow{\partial}_{J}\bigg){\cal O}_2;
\nn\\
\lbrack {\cal O}_1, {\cal O}_2\rbrack_*&\equiv& {\cal O}_1*{\cal O}_2-{\cal O}_2*{\cal O}_1.
\eea
The $\theta^{IJ}$ is the non-commutativity parameter, which is the inverse NS-NS field background \cite{Seiberg:1999vs}.
This non-commutative gauge theory can be mapped to the commutative gauge theory through the SW map \cite{Seiberg:1999vs}
$\hat{A}(A+\delta_{\lambda} A)=\hat{A}(A)+\hat{\delta}_{\hat{\lambda}}\hat{A}$, where 
\bea
\delta_{\lambda} A_{I}&\equiv&\partial_{I}\lambda; \qquad \hat{\delta}_{\hat{\lambda}}\hat{A}_{I}\equiv\partial_{I}\hat{\lambda}-\lbrack\hat{\lambda}, \hat{A}_{I}\rbrack_*.
\eea 
The solution of the SW map up to the first-order in the $\theta$ is given by \cite{Seiberg:1999vs}: 
\bea
\hat{A}_{I}=A_{I}-\theta^{JK}\bigg(A_{J}\partial_{K}A_{I}-\frac{1}{2}A_{J}\partial_{I}A_{K}\bigg)+{\cal O}(\theta^2); \qquad 
\hat{\lambda}=\lambda+\frac{1}{2}\theta^{IJ}A_{J}\partial_{I}\lambda+{\cal O}(\theta^2). 
\nn\\
\eea
After the SW map, the action for the commutative D3-brane theory in the NS-NS field background is obtained:
\bea
&&S_{CNSNS3}
\nn\\
&=&\int d^2xd^2\dot{x}\ \bigg(-\frac{1}{4}F^{IJ}F_{IJ}
+\frac{1}{2}F^{IJ}F_{IK}\theta^{KL}F_{LJ}
-\frac{1}{8}F^{IJ}F_{LK}\theta^{KL}F_{IJ}\bigg)
\nn\\
&=&
S_{NNSNS3}+{\cal O}(\theta^2),
\eea
which is up to a total derivative term. 
When we apply the large NS-NS field background limit or the scaling limit: 
\bea
l_s\sim\epsilon^{1/4};\ g_s\sim\epsilon^{1/2};\ B_{\dot{\mu}\dot{\nu}}\sim \epsilon^0; \ g_{\alpha\beta}\sim\epsilon^0; \ g_{\dot{\mu}\dot{\nu}}\sim\epsilon,
\eea 
the only non-vanishing component of the NS-NS field background is $B_{\dot{1}\dot{2}}$. 
In other words, only the non-commutativity parameter $\theta^{\dot{1}\dot{2}}$ remains. 
Note that the scaling limit of the NS-NS field background can be obtained from the scaling limit of the R-R field background through the S-duality \cite{Ho:2013opa}: 
\bea
l_s\sqrt{g_s}\rightarrow l_s; \ \frac{1}{g_s}\rightarrow g_s. 
\eea
Hence the commutative D3-brane theory in the large NS-NS field background is:
\bea
&&S_{CNSNS}
\nn\\
&=&\int d^2xd^2\dot{x}\ \bigg\lbrack-\frac{1}{4}F^{IJ}F_{IJ}
\nn\\
&&+g\bigg(-\frac{1}{2}F_{\dot{1}\dot{2}}F_{\dot{1}\dot{2}}F_{\dot{1}\dot{2}}
-\frac{1}{2}F_{\dot{1}\dot{2}}F_{01}F_{01}
-\frac{1}{2}F_{\dot{1}\dot{2}}F^{\alpha\dot{\mu}}F_{\alpha\dot{\mu}}
\nn\\
&&
-\frac{1}{2}\epsilon_{\alpha\beta\dot{\mu}\dot{\nu}}F^{\alpha\dot{\mu}}F^{\beta\dot{\nu}}F_{01}\bigg)\bigg\rbrack,
\eea
in which $g$ becomes $1/B_{\dot{1}\dot{2}}$.
After we replace $F$ by $G$ in the action $S_{CNSNS}$, we obtain $S_{EM}$. 
Hence we show that the non-commutative D3-brane theory in the large R-R field background is dual to the commutative D3-brane theory in the large NS-NS field background. 
Combing our result and the result of Ref. \cite{Ho:2015mfa} shows that the EM duality and SW map form a commutative diagram. 
This result implies that the solution of the SW map is consistent with the non-trivial test, S-duality. 
Later we apply the solution of the SW map to the multiple D4-branes. 
We also discuss the relation between the commutative and non-commutative descriptions.

\section{Discussion on Multiple D-branes}
\label{sec:4}
\noindent
By replacing a U(1) gauge group with a U($N$) gauge group, the single D-brane extends to the multiple D-branes in the large NS-NS field background.
The physical degrees of freedom are the same between the D3-brane theory in the R-R field background and the D3-brane in the NS-NS field background by the EM duality. 
Hence the multiple D-branes in the large R-R field background should be obtained in the same way as in the NS-NS field background case. 
The Yang-Mills theory should be the direct proposal for the non-commutative multiple D-branes in the large R-R field background \cite{Ho:2011yr}. 
The multiple D4-branes theory in the large R-R field background is
\bea
&&S_{NMRR4CF}
\nn\\
&=&
\int d^2xd^3\dot{x}\ \bigg\lbrack-\frac{1}{12}\hat{{\cal H}}_{\dot{1}\dot{2}\dot{3}}\hat{{\cal H}}^{\dot{1}\dot{2}\dot{3}}
+\frac{1}{2g}\epsilon^{\alpha\beta}\hat{{\cal F}}^{\mathrm{U}(1)}_{\alpha\beta}
-\frac{1}{4}\hat{{\cal F}}_{\dot{\nu}\dot{\rho}}^{\mathrm{U}(1)}\hat{{\cal F}}^{\dot{\nu}\dot{\rho}, \mathrm{U}(1)}
+\frac{1}{2}\hat{{\cal F}}_{\alpha\dot{\mu}}^{U(1)}\hat{{\cal F}}^{\alpha\dot{\mu}, U(1)}
\nn\\
&&
-\frac{1}{4}\mathrm{Str}\bigg(\hat{{\cal F}}_{IJ}^{\mathrm{SU}(N)}\hat{{\cal F}}^{IJ, \mathrm{SU}(N)}\bigg)\bigg\rbrack,
\eea
where 
\bea
\mathrm{Str}({\cal O}_1{\cal O}_2\cdots{\cal O}_n)&\equiv&
\mathrm{Tr}\big(\mathrm{Sym}({\cal O}_1{\cal O}_2\cdots{\cal O}_n)\big); 
\nn\\
\mathrm{Sym}({\cal O}_1{\cal O}_2\cdots{\cal O}_n)&\equiv&\frac{1}{n!}({\cal O}_1{\cal O}_2\cdots {\cal O}_n+ \mathrm{all\ permutations}).
\eea 
The trace operation is in the adjoint representation. 
Here we choose the symmetrized trace as in the NS-NS field background \cite{Tseytlin:1997csa}.
In this construction, the U($N$) gauge potential $a$ is already decomposed by the U(1) gauge potential $a^{\mathrm{U}(1)}$ and the SU($N$) gauge potential $a^{\mathrm{SU}(N)}$. 
The SU($N$) generator satisfies the properties: 
\bea
\mathrm{Tr}(T^a)=0; \qquad \mathrm{Tr}(T^aT^b)=N\delta^{ab}; \qquad \mathrm{Tr}(T^aT^bT^c)=\frac{i}{2}Nf^{abc},
\eea
 where $f^{abc}$ is an anti-symmetric tensor.
For the pure U(1) sector, it is trivially the same as the single D-brane theory. 
The non-trivial part is the SU($N$) Yang-Mills term. 
The gauge potential $b$ offers the non-trivial interaction between the $a^{\mathrm{U}(1)}$ and $a^{\mathrm{SU}(N)}$ in the SU($N$) Yang-Mills term. 
Therefore, we want to use the SW map to examine whether the proposal is the same as replacing the gauge group
\bea
&&S_{CMRR4a}
\nn\\
&=&\int d^2xd^3\dot{x}\ \mathrm{Str}\bigg\lbrack-\frac{1}{4}F_{IJ}F^{IJ}
\nn\\
&&+g\bigg(
\frac{1}{2}F_{01}F_{01}F_{01}+\frac{1}{4}F_{01}F^{\dot{\mu}\dot{\nu}}F_{\dot{\mu}\dot{\nu}}
-\frac{1}{2}F_{01}F^{\alpha\dot{\mu}}F_{\alpha\dot{\mu}}
-\frac{1}{2}\epsilon_{\alpha\beta}F^{\alpha\dot{\mu}}F^{\beta\dot{\nu}}F_{\dot{\mu}\dot{\nu}}
\bigg)\bigg\rbrack.
\nn\\
\eea  
\\

\noindent
The solution of the SW map up to the first-order gives: 
\bea
F_{\dot{\mu}\dot{\nu}}&=&\hat{F}_{\dot{\mu}\dot{\nu}}
-g(\partial_{\dot{\mu}}\hat{b}^{\dot{\rho}}\hat{F}_{\dot{\rho}\dot{\nu}}
+\hat{b}^{\dot{\rho}}\partial_{\dot{\rho}}\hat{F}_{\dot{\mu}\dot{\nu}}
+\partial_{\dot{\nu}}\hat{b}^{\dot{\rho}}\hat{F}_{\dot{\mu}\dot{\rho}})
+{\cal O}(g^2);
\nn\\ 
F_{\alpha\dot{\mu}}&=&\hat{F}_{\alpha\dot{\mu}}
-g(\partial_{\alpha}\hat{b}^{\dot{\rho}}\hat{F}_{\dot{\rho}\dot{\mu}}
+\hat{b}^{\dot{\rho}}\partial_{\dot{\rho}}\hat{F}_{\alpha\dot{\mu}}
+\partial_{\dot{\mu}}\hat{b}^{\dot{\rho}}\hat{F}_{\alpha\dot{\rho}})
+{\cal O}(g^2);
\nn\\
F_{\alpha\beta}&=&\hat{F}_{\alpha\beta}
-g(\partial_{\alpha}\hat{b}^{\dot{\rho}}\hat{F}_{\dot{\rho}\beta}
+\hat{b}^{\dot{\rho}}\partial_{\dot{\rho}}\hat{F}_{\alpha\beta}
+\partial_{\beta}\hat{b}^{\dot{\rho}}\hat{F}_{\alpha\dot{\rho}})
+{\cal O}(g^2),
\eea 
in which $\hat{b}^{\dot{\mu}}=-\partial^{\dot{\mu}}\dot{\partial}^{-2}\hat{F}_{01}^{\mathrm{U}(1)}+{\cal O}(g)$, where $\dot{\partial}^2\equiv\partial_{\dot{\mu}}\partial^{\dot{\mu}}$. 
The covariant field strength for the SU($N$) gauge group is expanded up to the first-order in $g$ and gives:
\bea
&&
\hat{{\cal F}}_{\dot{\mu}\dot{\nu}}^{\mathrm{SU}(N)}
\nn\\
&=&\hat{F}_{\dot{\mu}\dot{\nu}}^{\mathrm{SU}(N)}
\nn\\
&&
+g\bigg(\partial_{\dot{\sigma}}\hat{b}^{\sigma}\hat{F}^{\mathrm{SU}(N)}_{\dot{\mu}\dot{\nu}}
-\partial_{\dot{\mu}}\hat{b}^{\dot{\sigma}}\hat{F}^{\mathrm{SU}(N)}_{\dot{\sigma}\dot{\nu}}
-\partial_{\dot{\nu}}\hat{b}^{\dot{\sigma}}\hat{F}^{\mathrm{SU}(N)}_{\dot{\mu}\dot{\sigma}}\bigg)
\nn\\
&&+
{\cal O}(g^2);
\nn\\
&&
\hat{{\cal F}}_{\alpha\dot{\mu}}^{\mathrm{SU}(N)}
\nn\\
&=&\hat{F}_{\alpha\dot{\mu}}^{\mathrm{SU}(N)}
\nn\\
&&
+g\bigg(\partial_{\dot{\mu}}\hat{b}^{\dot{\sigma}}\hat{F}_{\dot{\sigma}\alpha}^{\mathrm{SU}(N)}
+\partial_{\alpha}\hat{b}^{\dot{\sigma}}\hat{F}_{\dot{\mu}\dot{\sigma}}^{\mathrm{SU}(N)}
+\epsilon_{\alpha\gamma}\hat{F}^{\gamma\dot{\sigma}, \mathrm{U}(1)}\hat{F}_{\dot{\mu}\dot{\sigma}}^{\mathrm{SU}(N)}\bigg)
\nn\\
&&
+{\cal O}(g^2);
\nn\\
&&\hat{{\cal F}}_{\alpha\beta}^{\mathrm{SU}(N)}
\nn\\
&=&\hat{F}_{\alpha\beta}^{\mathrm{SU}(N)}
\nn\\
&&
+g\bigg\lbrack-\hat{F}_{\alpha\dot{\mu}}^{\mathrm{SU}(N)}\bigg(\partial_{\beta}\hat{b}^{\dot{\mu}}+\epsilon_{\beta\gamma}\hat{F}^{\gamma\dot{\mu}, \mathrm{U}(1)}\bigg)
-\hat{F}_{\dot{\mu}\beta}^{\mathrm{SU}(N)}\bigg(\partial_{\alpha}\hat{b}^{\dot{\mu}}+\epsilon_{\alpha\gamma}\hat{F}^{\gamma\dot{\mu}, \mathrm{U}(1)}\bigg)\bigg\rbrack
\nn\\
&&
+{\cal O}(g^2).
\nn\\
\eea
Hence the non-commutative version of the action $S_{CMRR4a}$ is given by
\bea
&&S_{NMRR4a}
\nn\\
&=&\int d^2xd^3\dot{x}\ \mathrm{Str}\bigg\lbrack-\frac{1}{4}\hat{F}_{IJ}\hat{F}^{IJ}
\nn\\
&&+g\bigg(
\hat{F}_{01}^{\mathrm{U}(1)}\hat{F}_{01}^{\mathrm{SU}(N)}\hat{F}^{\mathrm{SU}(N)}_{01}
\nn\\
&&
+\frac{1}{2}\hat{F}^{\mathrm{U}(1)}_{01}\hat{F}^{\dot{\mu}\dot{\nu}, \mathrm{U}(1)}\hat{F}_{\dot{\mu}\dot{\nu}}^{\mathrm{U}(1)}
\nn\\
&&
+\frac{1}{2}\hat{F}_{01}^{\mathrm{U}(1)}\hat{F}^{\dot{\mu}\dot{\nu}, \mathrm{SU}(N)}\hat{F}_{\dot{\mu}\dot{\nu}}^{\mathrm{SU}(N)}
+\frac{1}{2}\hat{F}_{01}^{\mathrm{SU}(N)}\hat{F}^{\dot{\mu}\dot{\nu}, \mathrm{SU}(N)}\hat{F}_{\dot{\mu}\dot{\nu}}^{\mathrm{U}(1)}
\nn\\
&&
-\hat{F}_{01}^{\mathrm{SU}(N)}\hat{F}^{\alpha\dot{\mu}, \mathrm{SU}(N)}\hat{F}_{\alpha\dot{\mu}}^{\mathrm{U}(1)}
\nn\\
&&
-\frac{1}{2}\epsilon_{\alpha\beta}\hat{F}^{\alpha\dot{\mu}}\hat{F}^{\beta\dot{\nu}}\hat{F}_{\dot{\mu}\dot{\nu}}
\nn\\
&&
+\partial_{\dot{\mu}}\hat{b}^{\dot{\rho}}\hat{F}^{\dot{\mu}\dot{\nu}}\hat{F}_{\dot{\rho}\dot{\nu}}
+\partial_{\alpha}\hat{b}^{\dot{\rho}}\hat{F}^{\alpha\dot{\mu}}\hat{F}_{\dot{\rho}\dot{\mu}}
+\partial_{\dot{\mu}}\hat{b}^{\dot{\rho}}\hat{F}^{\alpha\dot{\mu}}\hat{F}_{\alpha\dot{\rho}}
-\epsilon^{\alpha\beta}\partial_{\alpha}\hat{b}^{\dot{\rho}}\hat{F}_{01}\hat{F}_{\dot{\rho}\beta}
\bigg)\bigg\rbrack.
\nn\\
\eea  
The SU($N$) Yang-Mills term gives
\bea
&&
-\frac{1}{4}\int d^2xd^3\dot{x}\ \mathrm{Str}\bigg(\hat{\cal F}_{\mu\nu}^{\mathrm{SU}(N)}\hat{{\cal F}}^{\mu\nu, \mathrm{SU}(N)}\bigg)
\nn\\
&=&-\frac{1}{4}\int d^2xd^3\dot{x}\ \mathrm{Str}\bigg(\hat{F}_{\mu\nu}^{\mathrm{SU}(N)}\hat{F}^{\mu\nu, \mathrm{SU}(N)}\bigg)
\nn\\
&&+g\int d^2xd^3\dot{x}\ \mathrm{Str}\bigg(
\frac{1}{2}\hat{F}_{01}^{\mathrm{U}(1)}\hat{F}^{\dot{\mu}\dot{\nu}, \mathrm{SU}(N)}\hat{F}_{\dot{\mu}\dot{\nu}}^{\mathrm{SU}(N)}
+\partial_{\dot{\mu}}\hat{b}^{\dot{\sigma}}\hat{F}^{\dot{\mu}\dot{\nu}, \mathrm{SU}(N)}\hat{F}_{\dot{\sigma}\dot{\nu}}^{\mathrm{SU}(N)}
\nn\\
&&
-\partial_{\dot{\mu}}\hat{b}^{\dot{\sigma}}\hat{F}^{\alpha\dot{\mu}, \mathrm{SU}(N)}\hat{F}_{\dot{\sigma}\alpha}^{\mathrm{SU}(N)}
-\partial_{\alpha}\hat{b}^{\dot{\sigma}}\hat{F}^{\alpha\dot{\mu}, \mathrm{SU}(N)}\hat{F}_{\dot{\mu}\dot{\sigma}}^{\mathrm{SU}(N)}
\nn\\
&&
-\epsilon_{\alpha\gamma}\hat{F}^{\gamma\dot{\sigma}, \mathrm{U}(1)}\hat{F}^{\alpha\dot{\mu}, \mathrm{SU}(N)}
\hat{F}_{\dot{\mu}\dot{\sigma}}^{\mathrm{SU}(N)}
\nn\\
&&
-\epsilon^{\alpha\beta}\partial_{\beta}\hat{b}^{\dot{\mu}}\hat{F}_{01}^{\mathrm{SU}(N)}\hat{F}_{\alpha\dot{\mu}}^{\mathrm{SU}(N)}
-\hat{F}_{01}^{\mathrm{SU}(N)}\hat{F}^{\alpha\dot{\mu}, \mathrm{U}(1)}\hat{F}_{\alpha\dot{\mu}}^{\mathrm{SU}(N)}\bigg)+{\cal O}(g^2).
\eea
The result shows that the SU($N$) Yang-Mills theory loses the below terms
\bea
&&S_{NMRR4a}-S_{NMRR4CF}
\nn\\
&=&
g\int d^2xd^3\dot{x}\ \mathrm{Str}\bigg(\hat{F}_{01}^{\mathrm{U}(1)}\hat{F}_{01}^{\mathrm{SU}(N)}\hat{F}_{01}^{\mathrm{SU}(N)}
+\frac{1}{2}\hat{F}_{01}^{\mathrm{SU}(N)}\hat{F}^{\dot{\mu}\dot{\nu}, \mathrm{SU}(N)}\hat{F}_{\dot{\mu}\dot{\nu}}^{\mathrm{U}(1)}
\nn\\
&&
-\frac{1}{2}\epsilon_{\alpha\beta}\hat{F}^{\alpha\dot{\mu}, \mathrm{SU}(N)}\hat{F}^{\beta\dot{\nu}, \mathrm{SU}(N)}\hat{F}_{\dot{\mu}\dot{\nu}}^{\mathrm{U}(1)}\bigg)+{\cal O}(g^2).
\eea
Hence we show that the non-commutative multiple D4-branes theory in the large R-R field background is different from the proposal. 
Because the degrees of freedom for the gauge potential $b^{\dot{\mu}}$ field is dual to the gauge potential $a_{\alpha}$, and we know that the gauge group of all one-form gauge potentials is U($N$), the difference should be expected. 
The proposal of Ref. \cite{Ho:2011yr} should be wrong due to the inconsistency.

\section{Outlook}
\label{sec:5}
\noindent
Our result suggests that the non-Abelian field strength associated with the gauge potential $b$ is necessary for the large R-R field background. 
This result should shed light on the non-commutative multiple M5-brane theory. 
The construction of a large R-R field background breaks the Lorentz symmetry and leads to a property, non-locality \cite{Ho:2011yr}. 
The non-locality is generated by $\dot{\partial}^{-2}$ for the D-branes in the large R-R field background \cite{Ho:2011yr}. 
When one studies the low-momentum mode, the non-local term $\dot{\partial}^{-2}$ is divergent. 
This non-local term should be problematic, but it is not because we take the large R-R field background limit first and then the low-energy limit.  
The multiple M5-branes theory possibly does not have an action with the Lorentz symmetry. 
However, the large $C$-field background breaks the symmetry \cite{Ho:2008nn}. 
It should help the multiple M5-branes theory without the Lorentz symmetry in the following directions: (1) Extending the solution of SW map to all-orders; (2) Defining the Nambu-Posson bracket with the U($N$) gauge group; (3) Writing the action in terms of the covariant field strength.  

\section*{Acknowledgments}
\noindent
The author would like to thank Xing Huang and Yiwen Pan for their useful discussion and thank Nan-Peng Ma for his encouragement.
\\

\noindent
The author was supported by the Post-Doctoral International Exchange Program; 
China Postdoctoral Science Foundation, Postdoctoral General Funding: Second Class (Grant No. 2019M652926); 
Foreign Young Talents Program (Grant No. QN20200230017); 
Science and Technology Program of Guangzhou (Grant No. 2019050001).


\end{document}